# Persistence Characterisation of Teledyne H2RG detectors

Simon Tulloch

European Southern Observatory, Karl Schwarzschild Strasse 2, Garching, 85748, Germany.

**Abstract.** Image persistence is a major problem in infrared detectors, potentially seriously limiting data quality in many observational regimes. The problem manifests itself as remnant images that can persist for several days after a deep exposure. In this study, the persistence behavior of three 5.3um cutoff H2RGs has been characterised using a low-background cryostat with LED light sources. Persistence charge de-trapping was measured over several hours following a wide range of exposure levels and exposure times. This data was then analysed to yield charge trapping and de-trapping spectra which present graphically the trap density as a function of their time constants. These spectra show the detector behavior in a very direct way and offer a natural metric for comparing different devices. It is hoped that the trap time constant spectra for each detector can be used in an analysis pipeline to remove persistence artifacts based on the recent exposure history of the detector. The study confirmed that the charge traps responsible for persistence must be present in the depletion region of the pixel, however, two trap populations were revealed. One of these captures charge within milliseconds and then releases it over many hours. The second population is less problematic with fairly similar trapping and de-trapping time constants. Large differences in persistence magnitude and trap spectra have been found even between devices with near-consecutive serial numbers. Lower temperatures resulted in lower persistence both in terms of total trapped charge and the time taken for that charge to decay. Limiting the full-well by reducing pixel bias voltage also had a beneficial effect. Previously proposed mitigation techniques including "global reset de-trapping" and "night light" illumination were tried but found to be ineffective.

## 1 Trap Model of Persistence

The trap model[1] proposes that depletion region traps are the principle source of image persistence. These traps are filled with charge as the depletion region shrinks under exposure to light. They then emit charge once the depletion region is reset to its original width prior to the next exposure. The process is shown below in Figure 1.

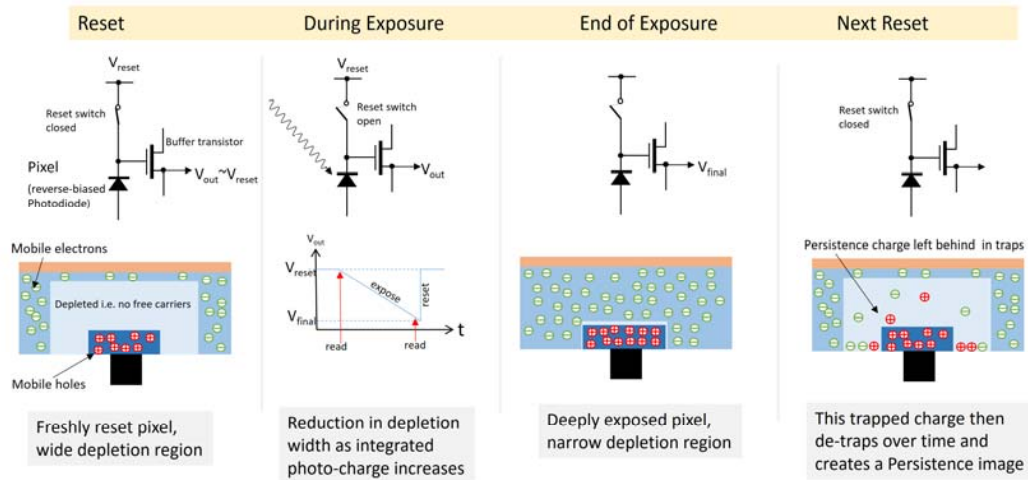

**Fig. 1.** The depletion region of a photodiode during the various phases of imaging, showing the trapping behavior.

## 2      Characterisation Technique

The persistence measurement process is shown in Figure 2. First the detector was allowed to  settle in the dark for several hours. The LED was then flashed to deliver a known amount of photo-charge. The pixels were then left in their exposed state for varying lengths of time before being reset. During this time the traps in the pixels will be "soaked" in the photo-charge and will capture some of it. At the end of the soak the pixels were reset and a series of 400 non-destructive reads taken over a period of 8000s. These frames were geometrically spaced to better sample the growth of the persistence image as the traps gradually released charge. Only a small region of the detector was illuminated  which allowed a shadowed comparison region to be used for dark current and glow subtraction. The measurements were performed on a mosaic of three nearly identical detectors.

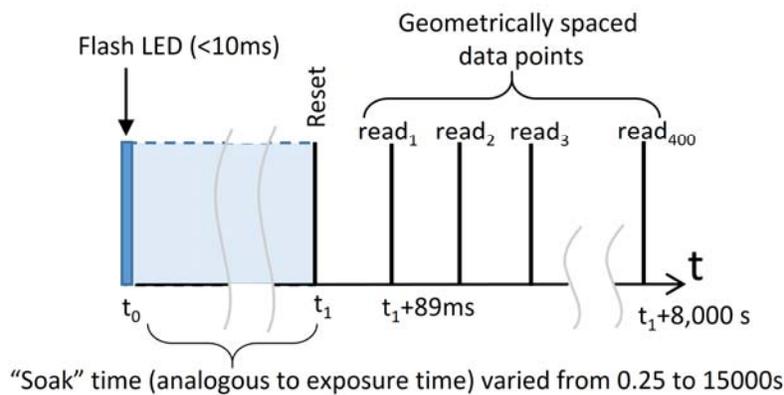

**Fig. 2.**Persistence measurement technique using brief illumination by an LED. The wavelength was 935nm.

## 3      Persistence versus signal level

The first measurement was of how persistence varies with exposure level while keeping the soak time constant (typically 30s). The on-time of the LED as adjusted to give a range of exposures up to a few times the full-well of the pixel. The persistence was found to increase with the exposure level up to full-well. Above full-well the persistence then got no worse. This is consistent with the depletion region trap model since once the depletion region was fully collapsed by a deep exposure its volume would not then change further with additional illumination.

## 4      Persistence versus exposure time

In this series of measurements the exposure given by the LED was kept constant at just below full-well and the soak time varied between 0 and 15000s. The result of 20 such runs is shown in Figure 3. The shorter soak times produced a de-trapped charge profile that quickly levelled off indicating short de-trapping times. Longer soak times gave correspondingly longer de-trapping times. These charge profiles were then differentiated to yield the persistence current and this is shown in the lower panel of the figure. The particular device shown in Figure 3 had a persistence current equal to the dark current $(0.01e^-/s)$ 1000s after a 60s exposure of 100ke$^-$.

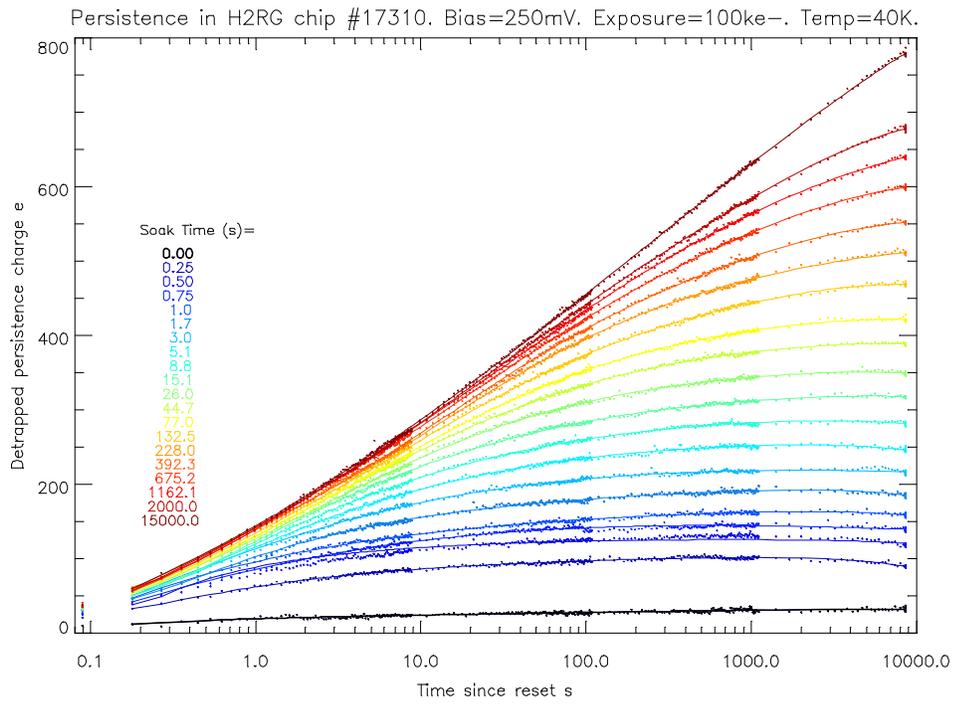

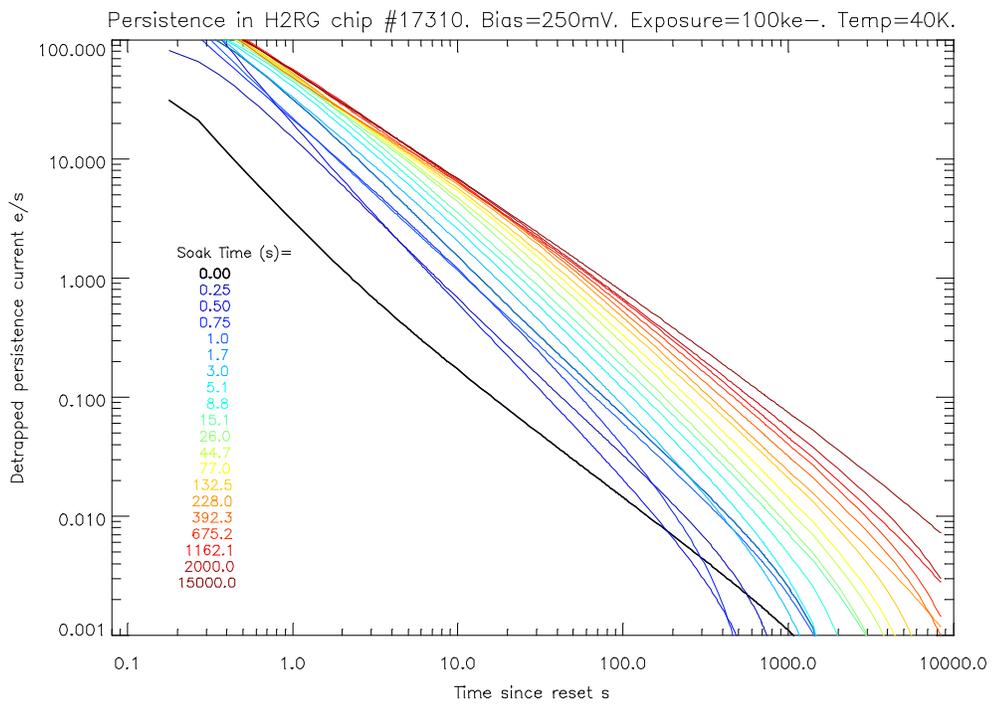

**Fig. 3.** Persistence charge and current profiles for a constant initial exposure (100ke⁻) and a wide range of soak times.

# 5    Persistence charge statistics

The persistence model[1]  suggests that de-trapped electrons and holes originating in depletion region traps will effectively have fractional charge. They will therefore give a different number of e⁻/ADU than normal photo-generated charge. This is somewhat counterintuitive but is argued convincingly from an energy-conservation angle. The idea was tested experimentally for all three detectors by using two identically exposed persistence image sequences. These were then subjected to the standard photon transfer analysis used to calculate system gain.  Special precautions had to be taken to prevent "hot" or dead pixels from disturbing the statistics. When measuring the variance, a histogram of the persistence-affected pixels in the difference image was fitted with a Gaussian function whose width was then used to calculate the variance. One of the two input images was similarly analysed with the centre point of the Gaussian used to give the mean. A second dark current window on each chip was analysed using the same technique as a check that the correct e⁻/ADU value was found for "conventionally" produced signal (dark charge and photo-charge have the same statistical properties). The histogram technique was very effective at rejecting faulty pixels from the analysis. The result is shown in Figure 4. The dark charge gave a gain of 2e⁻/ADU exactly as expected from other independent measurements. Very interestingly the persistence charge gave a higher e⁻/ADU result, consistent with an effective carrier charge of 0.4 e⁻.

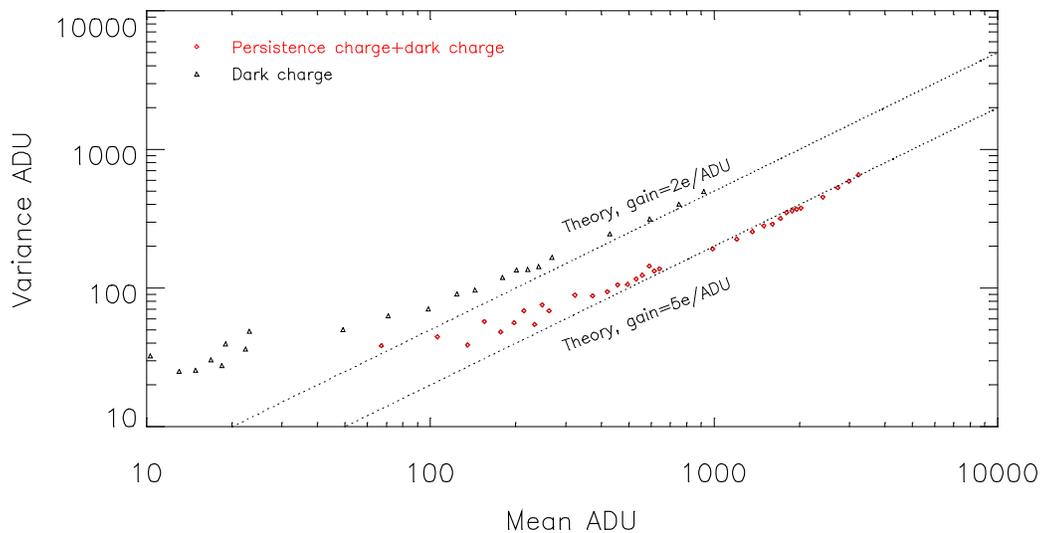

**Fig. 4.**Photon transfer analysis of persistence charge and dark charge.

# 6    Electrically generated persistence

The effect of optical illumination on the depletion region can be simulated electrically by holding the pixel in reset and reducing the reset voltage bias line  i.e. by reducing the reverse bias applied to the diode for the soak time and then returning it to its initial value at the end of the exposure. A subsequent long dark frame was used to capture the de-trapped charge and produce an electrically induced persistence map. This is shown in Figure 5. All three devices tested showed very similar performance. Circular structures were visible corresponding to regions of lower trap density. It was expected that a similar map produced by optical stimulation would be the same, however, this was not the case for device #17308. This can be seen in Figure 6.

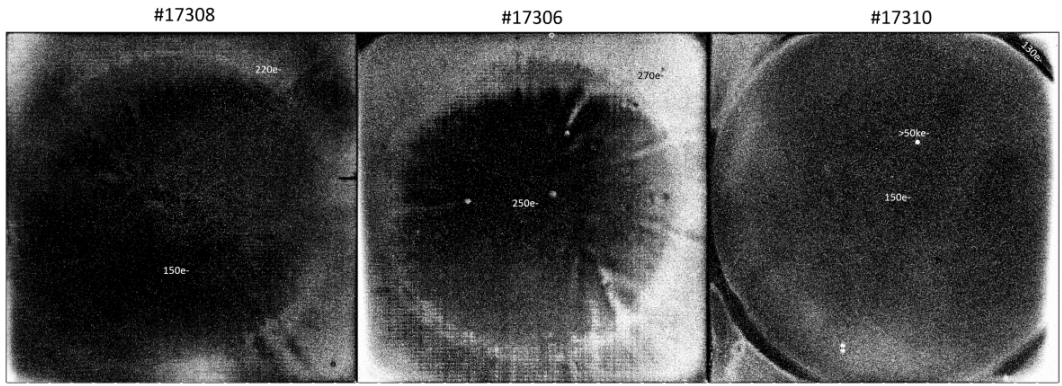

**Fig. 5.** Electrically induced persistence maps .

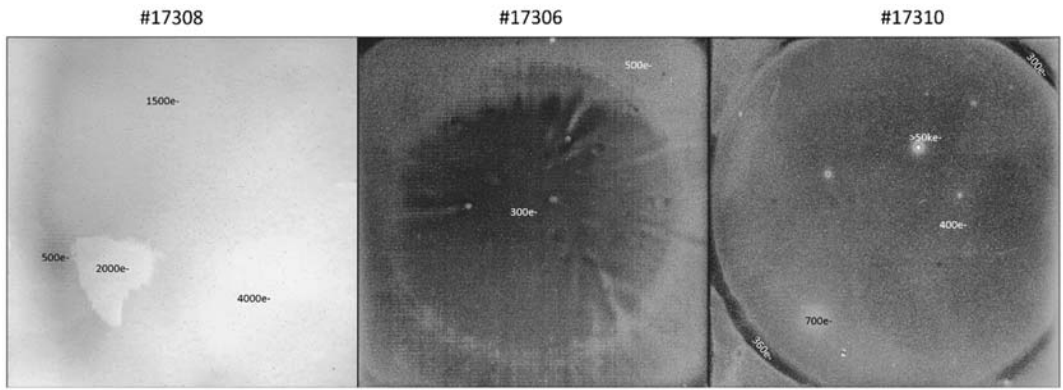

**Fig. 6.** Optically induced persistence maps.

## 7 Mitigation Techniques

### 7.1 Global reset de-trapping

This is the technique of holding a detector in an extended global reset or doing a large number of short resets in an attempt to accelerate charge de-trapping. It is entirely ineffective. Since global reset results in a slightly different reset voltage being applied to a pixel compared to that applied by a line by line reset it can in fact release additional persistence charge that may be mistaken for dark current.

### 7.2 Bias voltage reduction

Reducing the reverse bias on the pixel photodiode will reduce the full-well. It will also reduce the maximum volume of the depletion region thereby reducing the number of traps capable of generating persistence. To test this the full-well and persistence were measured at reverse biases of 0mV and 230mV. Persistence was measured by illuminating with a fluence of 94ke⁻, soaking for 30s and then recording a 100s dark frame. With a 0mV bias the full-well was found to be 50ke⁻ and the persistence was 110e⁻. For the default 230mV reverse bias case the full-well increased to 120ke⁻ and the persistence increased to 230e⁻. Tuning the full-well of the pixel to the observational regime could thus offer a small improvement in persistence performance.

### 7.3 Sub-division of frames

The model suggests that persistence will be proportional to the swept volume of the depletion region over the integration/reset cycle. The simplest way to reduce the swept volume for a given observation is to increase the frame rate. This idea was tested by reducing the LED light source flux so that a fluence of approximately half-full was reached in 100s. A single 100s exposure was taken, the LED switched off, the detector reset and the subsequent de-trapped charge measured using a sequence of non-destructive reads. Next, the same 100s exposure was repeated but with an extra reset after 50s thus effectively dividing the exposure into two sub-exposures with twice the frame rate. Further repetitions were done with up to 8 additional resets during the integration period. The method is illustrated in Figure 7.The persistence results are shown in Figure 8.

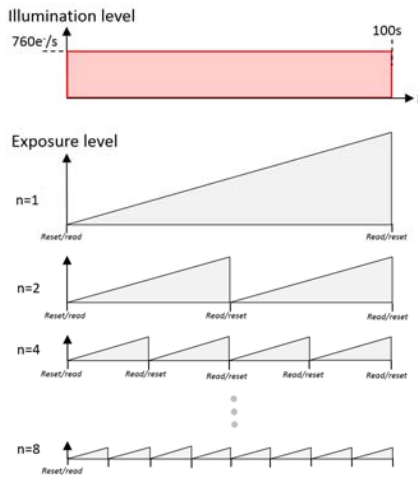

**Fig. 7.** Sub-division of a frame using higher frame rates and its effect on maximum exposure level.

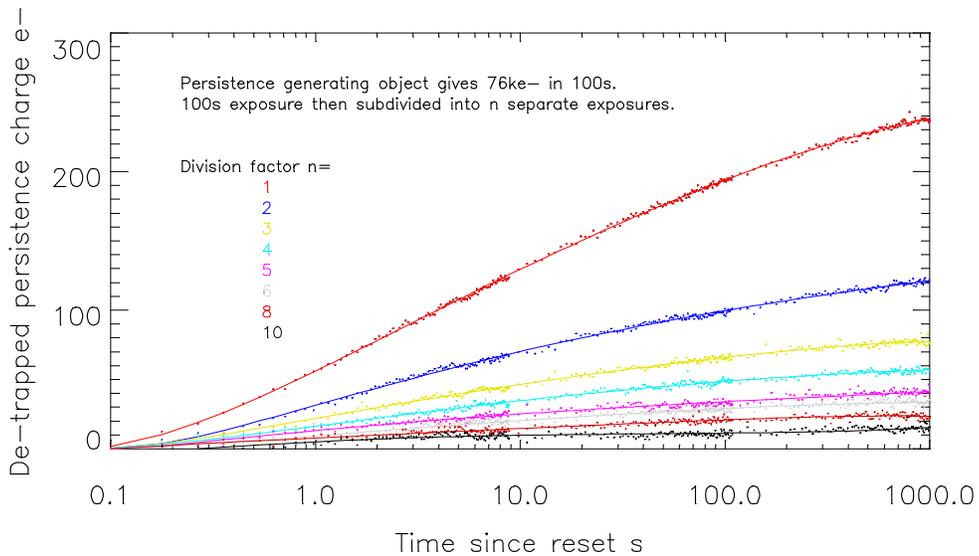

**Fig. 8.** Effect of frame sub-division on persistence charge de-trapped over 1000s. Chip #17310 @41K

### 7.4 Cooling

The temperature of the detectors in the test cryostatwas varied between 41K and 55K to see the effect on persistence. The results are shown for two detectors in Figures 9,10. Device #17308 (the one that displayed such varied response to electrical and optical stimulation) suffered a huge increase in persistence at the elevated temperatures.

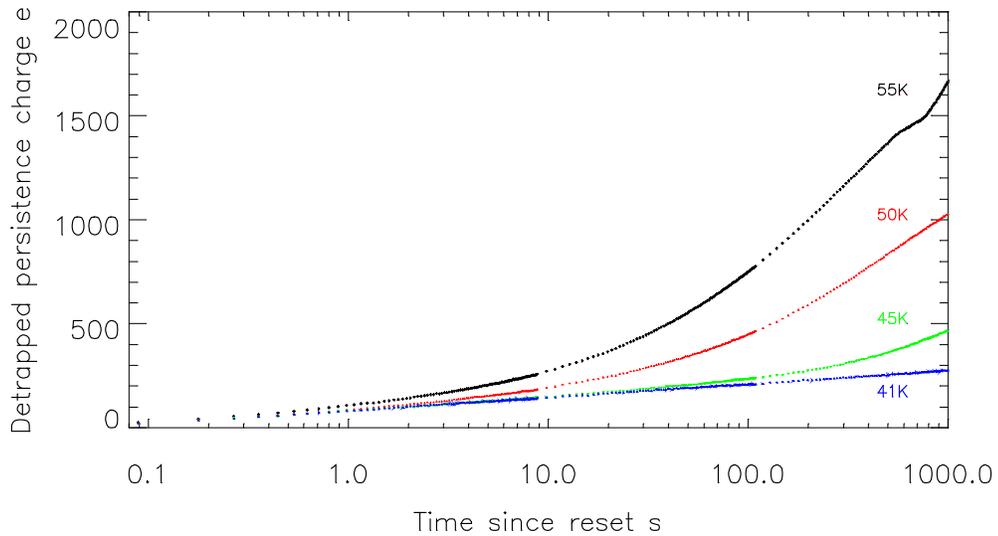

**Fig. 9.** Persistence in device #17308 following 74ke⁻ exposure and 100s soak at various temperatures.

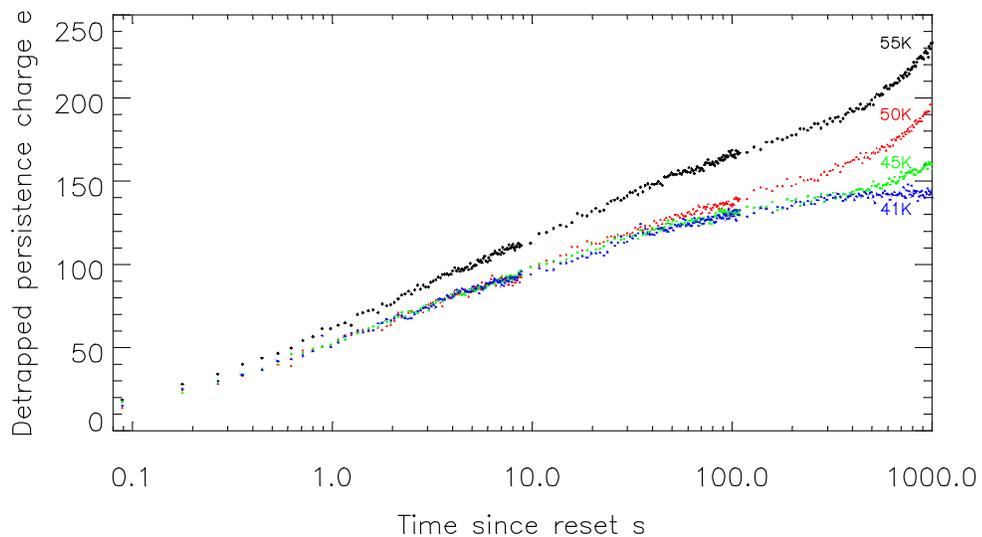

**Fig. 20.** Persistence in device #17310 following 47ke⁻ exposure and 100s soak at various temperatures.

### 7.5 "Night Light"

This idea was first proposed by Roger Smith. The idea is that it may be possible to dislodge trapped charge by heavy illumination with light just beyond the cut-off wavelength of the detector material. Whilst not providing enough energy to lift electrons into the conduction band it may provide sufficient energy to de-trap charge from mid-band states thus accelerating the decay of persistence. The idea was tested using a wideband black-body source at 250K filtered through a 6.5um long-pass filter. The detector was first pre-flashed with an LED to induce plenty of persistence. It was then exposed to the "night light" for 40s over a small region of the persistence-affected area. Unfortunately no difference was seen between the region illuminated by the night light and regions that remained in shadow. The night light fluence was estimated to be $10^9$ photons between the cut-on wavelength of the filter and twice the cut-off wavelength of the detector. This is, however, an intriguing idea and it would be interesting to repeat the measurement with higher fluence and a filter edge placed closer to the cut-off wavelength.

## 8    Defining a Metric for persistence data

### 8.1    Overly simplistic methods

Since persistence is dependent on a combination of factors such as exposure level and exposure time it is very difficult to describe it simply with a single number as is the case with, for example, detector dark current or detector read noise. Setting requirements based on the maximum permitted persistence current at say 100s after a full-well exposure is meaningless if there are a large variety of de-trapping time constants present. This is illustrated below in Figure 11 where the persistence current of the three detectors is plotted as a function of time since reset for identical exposure levels and for soak times of 5 and 132s. Once again device #17308 stands out as being very badly behaved with some very long-time constant traps giving problematic persistence many hours after reset. However, the same chip at 100s was actually the best of the three at the longer soak time.

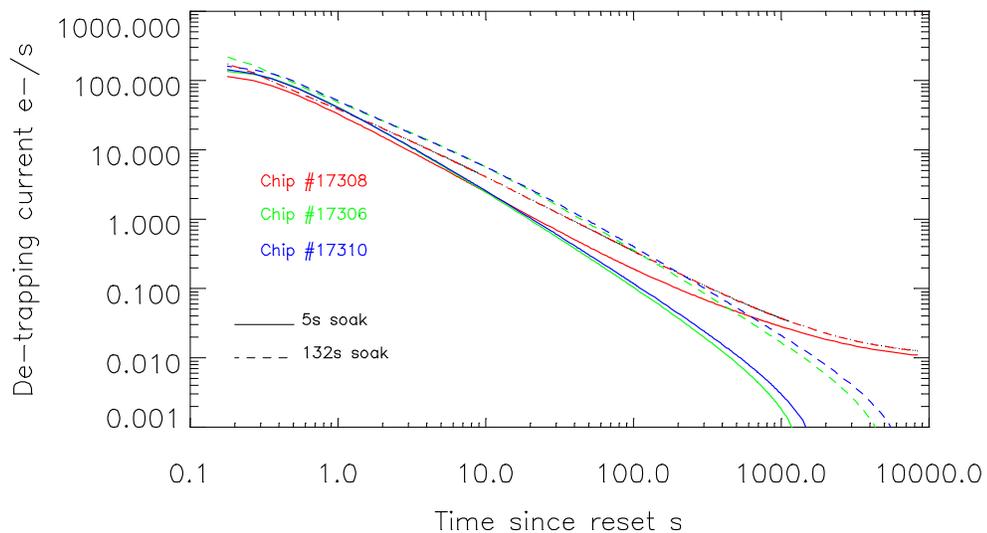

**Fig. 31.** Persistence current for all three detectors plotted together for two separate soak times.

## 8.2    Time constant spectra

A more complete description of persistence is needed and it must contain information on both the amount of charge that is trapped and the rapidity with which it is captured and emitted. The "Tau Spectra" metric is proposed here as a more optimal way of comparing persistence behavior of competing detectors. This new metric is described as follows.

First it is necessary to replot the persistence charge profiles over a range of soak times (essentially what is shown in the upper panel of Figure 3.) as a 3D "pyramid".  This is shown below in Figure 12. Edge 1 of this pyramid corresponds to very long soak times where we can be reasonably confident that most of the traps have been fully charged. Edge 1 then describes how this trapped charge decays over approximately 8000s (ideally it should have been even longer in the case of this particular detector). Edge 1 therefore contains information on the *de-trapping* time constants of the detector. Edge 2, on the other hand, corresponds to very long de-trapping times where we can confident that most of the traps have fully discharged. This edge therefore contains information on *trapping* time constants. If traps charged up and down at the same rate then Edges 1 and 2 would be identical, however, this was not the case.

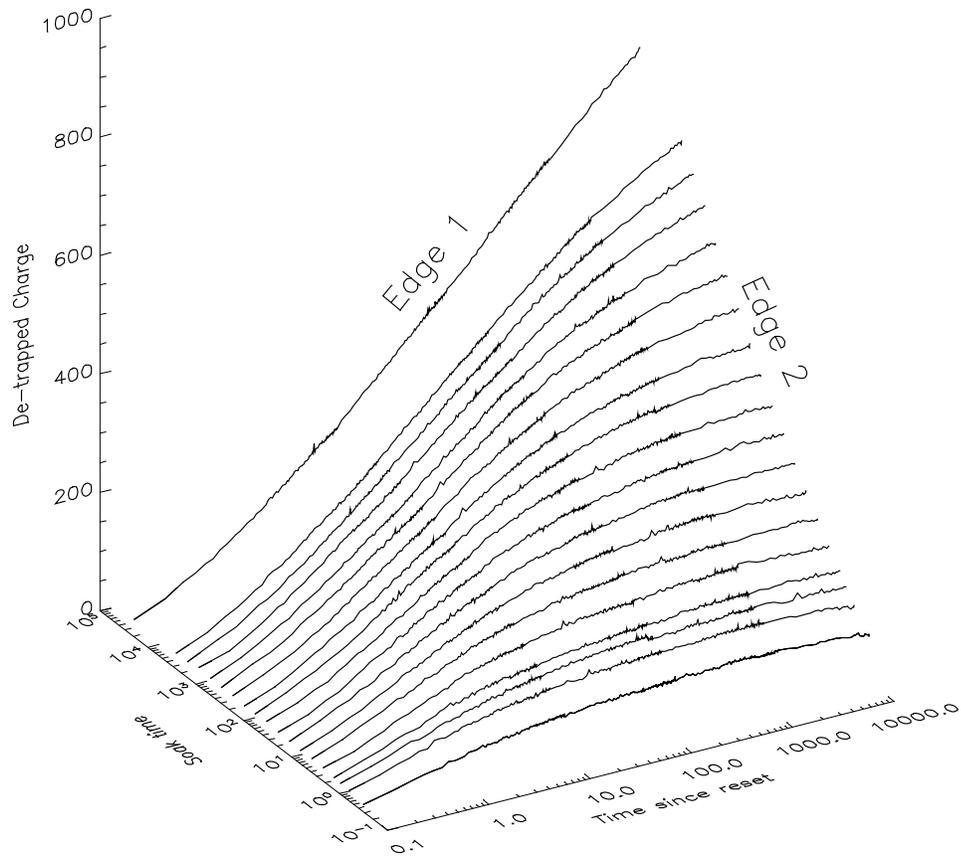

**Fig. 42.** De-trapped charge as a function of soak time and time since reset. The pyramid edges are labelled.

The two edges were then analysed in further detail. It was assumed that traps charged and discharged in a similar manner to an RC network i.e. with the following exponential function:

$$q(t) = A.\left\{1 - e^{(-t/\tau)}\right\},\qquad(1)$$

where :

$q(t) =$ trapped charge as a function of time (analogous to charge on capacitor),

$A =$ trap density,

$\tau =$ time constant of traps (analogous to an RC constant).

It was further assumed that the actual traps displayed a wide range of time constants and that the observed trapping profiles could be analysed as the sum of an ensemble of the functions shown in Equation 1. The two edges of the pyramid were then fitted with the following ensembles:

$$q(t) = \sum_{n=1}^{n=14} A_n\,(\tau_n).\left(1 - e^{t/\tau_n}\right)\qquad(2)$$

The result of one such fit is shown in Figure 13. Here an ensemble of 14 exponential functions gave an extremely good fit. The time constants of each function were geometrically spaced at half decade intervals.

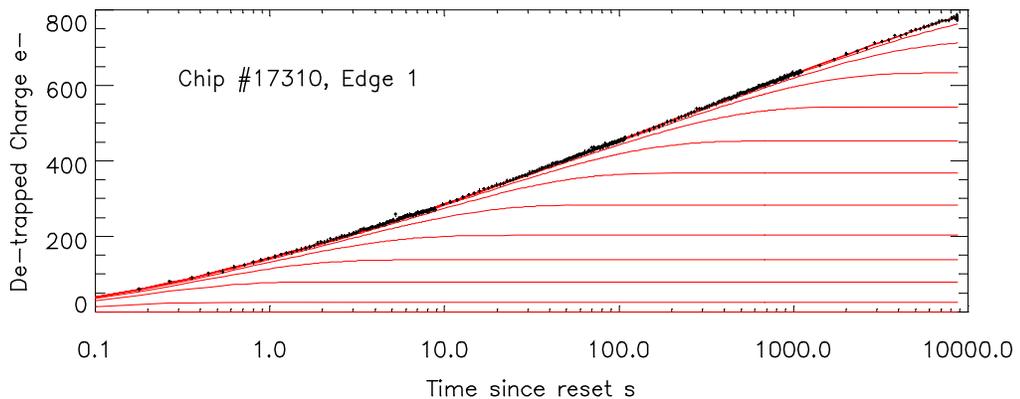

**Fig. 53.** The shape of Edge 1 in Figure 12 analysed as an ensemble of exponential functions.

From our fit we now have 14 values for $A_n$ that physically represent the amount of charge emitted into a pixel in each of our half-decade wide detrapping-time constant bins. Since the Edge 1 data corresponds to a very long soak time, this emitted charge will essentially be the trap density since most traps will have had time to fully charge up prior to the measurement. Plotting $A_n$ as a function of $\tau_n$ then gives us a "Tau spectrum". In the case of the Edge 1 data, $\tau_n$ actually represents the *de-trapping* time constants.

We now perform a similar analysis on Edge 2. This is shown in Fig 14. The Edge 2 data shows the amount of charge emitted by traps as a function of the initial soak time. Since this edge corresponds to very long de-trapping times (8000s) this emitted charge will essentially be the same as the amount of charge that was initially trapped. Each bin in the Tau spectra obtained from Edge 2 thus represents the trap density as a function of the *trapping* time constant.

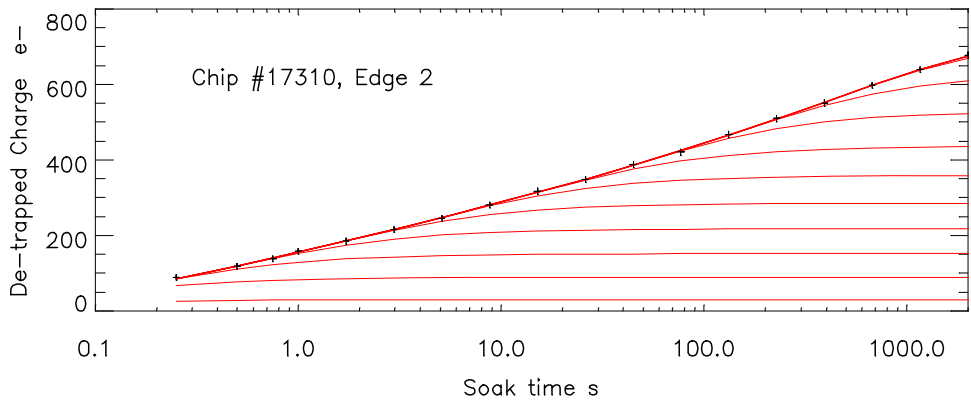

**Fig. 64.** The shape of Edge 2 in Figure 12 analysed as an ensemble of exponential functions.

The final result is shown in Figure 15. Here the y-axis is trap density per pixel for an initial exposure of 100ke⁻. We can see from this plot that the time constants of the traps are widely spread and that de-trapping is slower than trapping.

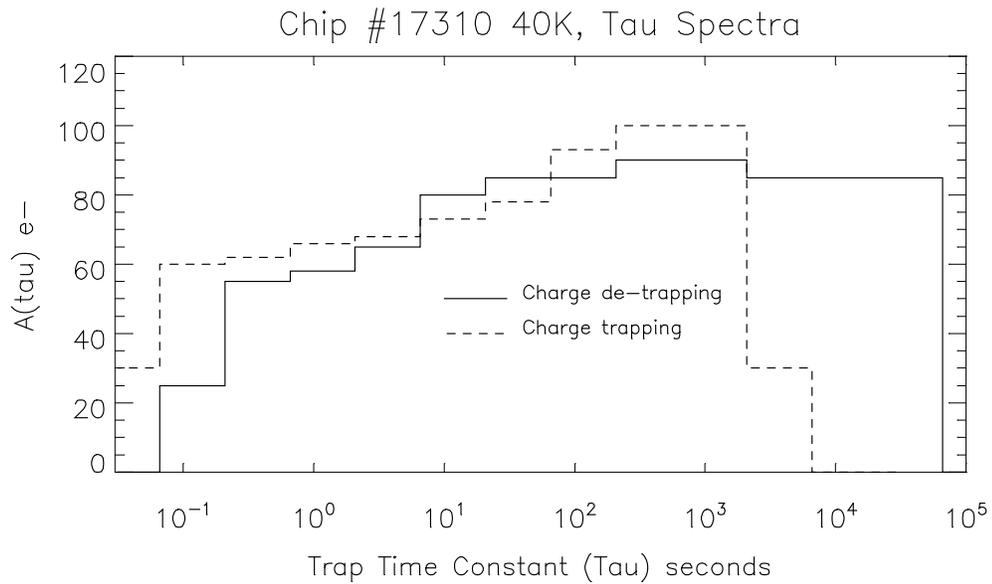

**Fig. 75.** The Tau spectra for device #17310. Data from both edges is shown overlaid.

A tau spectrum for detector #17308 was also obtained. Recall that this detector had already shown unusual behavior : it showed persistence that was long lived and highly temperature sensitive and responded differently to electrical and optical stimulation. Its spectrum is shown in Figure 16 and is notably different from that of #17310. The spectra show us that the device traps a large proportion of its persistence charge on timescales of less than 100ms. Unfortunately a large proportion of the total trapped charge is then released on a timescale of greater than $3 \times 10^4$s which makes this device particularly problematic in astronomical applications.

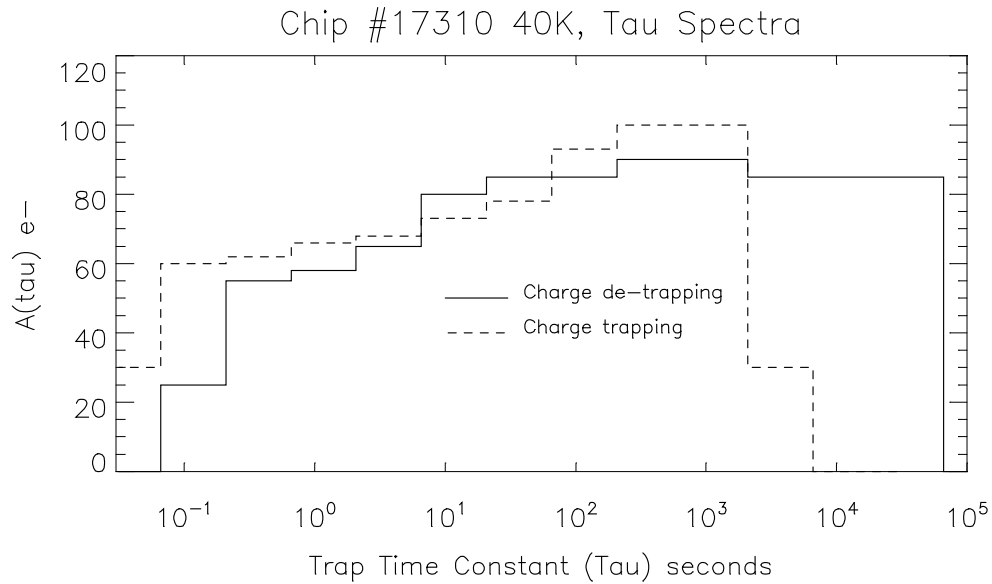

**Fig. 86.** The Tau spectra for device #17308. Initial exposure level was 100ke⁻.

## 9    Conclusions and further work

The third chip in the mosaic that was studied had serial number #17306. Its behavior was very similar to that of #17310. Device #17308 behaves so differently from these other two that it seems reasonable to assume that it contains a significant proportion of traps belonging to a second population. This second population perhaps corresponds to traps in a physically distinct part of the pixel. One hypothesis is that one population corresponds to surface traps and the other to bulk traps. The behavior of the surface traps will be dependent on the passivation used so it will be interesting to see the Tau spectrum of Tele-dyne's newer passivation process.

The Tau Spectra technique seems to be a very effective way of presenting graphically the persistence performance of any given detector. It allows easy comparison between devices.

Further work is needed to investigate trapping at shorter timescales, the susceptibility of H2RG detectors to persistence even when held in reset during illumination and the relationship between persis-tence and reciprocity failure.